\documentclass[aps,preprint,showpacs,floatfix]{revtex4}
\begin{document}
%\input psfig

%\preprint{{DOE/ER/XXXXX-XXX}\cr{UMPP\#05-XXX}}

\count255=\time\divide\count255 by 60 \xdef\hourmin{\number\count255}
  \multiply\count255 by-60\advance\count255 by\time
 \xdef\hourmin{\hourmin:\ifnum\count255<10 0\fi\the\count255}

\newcommand{\xbf}[1]{\mbox{\boldmath $ #1 $}}

\newcommand{\sixj}[6]{\mbox{$\left\{ \begin{array}{ccc} {#1} & {#2} &
{#3} \\ {#4} & {#5} & {#6} \end{array} \right\}$}}

\newcommand{\threej}[6]{\mbox{$\left( \begin{array}{ccc} {#1} & {#2} &
{#3} \\ {#4} & {#5} & {#6} \end{array} \right)$}}

\newcommand{\clebsch}[6]{\mbox{$\left( \begin{array}{cc|c} {#1} & {#2} &
{#3} \\ {#4} & {#5} & {#6} \end{array} \right)$}}

\newcommand{\iso}[6]{\mbox{$\left( \begin{array}{cc||c} {#1} & {#2} &
{#3} \\ {#4} & {#5} & {#6} \end{array} \right)$}}

\title{The Large $N_c$ Baryon-Meson $I_t \! = \! J_t$ Rule Holds for
Three Flavors}

\author{Richard F. Lebed}
\email{Richard.Lebed@asu.edu}

\affiliation{Department of Physics and Astronomy, Arizona State
University, Tempe, AZ 85287-1504}

%\date{\hourmin, \today}
\date{March, 2006}

\begin{abstract}
It has long been known that nonstrange baryon-meson scattering in the
$1/N_c$ expansion of QCD greatly simplifies when expressed in terms of
$t$-channel exchanges: The leading-order amplitudes satisfy the
selection rule $I_t \!  = \! J_t$.  We show that $I_t \!  = \! J_t$,
as well as $Y_t \! = \! 0$, also hold for the leading amplitudes when
the baryon and/or meson contain strange quarks, and also characterize
their $1/N_c$ corrections, thus opening a new front in the
phenomenological study of baryon-meson scattering and baryon
resonances.
\end{abstract}

%11.15.Pg   Expansions for large numbers of components (e.g.,
%           1/Nc expansions)
%14.20.Gk   Baryon resonances with S=0
%14.20.Jn   Hyperons

\maketitle

\section{Introduction} \label{intro}

A classic series of papers by Adkins, Nappi, and Witten~\cite{ANW}
written over two decades ago showed that a number of relations among
baryon observables in chiral soliton models, particularly the Skyrme
model, appear to be model independent and related to the large $N_c$
limit of QCD.  An extensive body of literature followed, notably
including work by the Siegen group and Mattis and
Karliner~\cite{HEHW,MK}, finding linear relations between $\pi N$
scattering amplitudes in various isospin and angular momentum channels
that hold at all energies.  Developing this theme, Mattis and
Peskin~\cite{MP} found a remarkable group structure to be responsible
for the relations: In the soliton language, the conserved underlying
quantum number in $s$-channel scattering is the ``grand spin'' $K$,
where ${\bf K} \! = \! {\bf I} \! + \! {\bf J}$.  Multiple observable
scattering amplitudes arise as linear combinations of a smaller set of
``reduced'' scattering amplitudes labeled by $K$.

Donohue subsequently noted~\cite{Donohue} the remarkable result that
these linear relations simplify dramatically when expressed in the $t$
channel.  Armed with this observation and SU(2) group theory
identities to manifest the crossing symmetry, Mattis and Mukerjee (MM)
proved underlying $K$-spin conservation for the 2-flavor system to be
equivalent to the rule $I_t \!  = \! J_t$ at large $N_c$~\cite{MMSU2}.
However, when attempting to extend their rule to include strangeness,
MM obtained~\cite{MMSU2,MMSU3} results not only that did not appear to
support the $I_t \! = \! J_t$ rule, but moreover that they could not
show to contain the 2-flavor results as a special case.  Even so,
using a 3-flavor Skyrme picture Donohue later
found~\cite{Donohue3flavor} that the number of independent amplitudes
reduces substantially when the problem is expressed in the $t$
channel.

The problem then lay largely unnoticed until a few years ago, when
Cohen and the present author revived the amplitude relation
approach~\cite{CL1st,CLcompat,CLconfig,CL1N,CLphoto,CLpenta,CLSU3,
CLSU3penta,CL70} in order to study baryon resonances in the $1/N_c$
expansion.  In addition to a number of successes, such as an
understanding of the large $N(1535)$ $\eta N$ coupling~\cite{CL1st}
and the large $N_c$ reason why the quark model produces a good but not
perfect accounting of the resonance spectrum~\cite{CLcompat}, we
considered more formal issues as well.  In particular, we noted that
the SU(3) group theory for baryons at arbitrary $N_c$ required
development both mathematically and in terms of the proper
identification of baryon quantum numbers~\cite{CLSU3,CLSU3penta,CL70}
and showed that the MM 3-flavor $s$-channel expression does indeed
reduce to the appropriate 2-flavor one~\cite{CL70}.

In this paper we show that, in fact, the $I_t \! = \! J_t$ rule holds
for the 3-flavor case as well, and also obtain a new selection rule
$Y_t \! = \! 0$.  We give a proof both using linear amplitude
expressions similar to those obtained by MM, and also using the more
recent operator approach~\cite{operator,KapSavMan}.  We further argue
that corrections to the $I_t \! = \! J_t$ rule among processes of a
given fixed strangeness go as $1/N_c^n$ for $|I_t \! - \! J_t| \!  =
\!  n$, while corrections to the $Y_t \! = \! 0$ only fall off as
$N_c^{-|Y_t|/2}$.

In Sec.~\ref{Tamp} we rederive the MM $t$-channel master scattering
expression, making appropriate corrections and modifications.  The
proofs of the $Y_t \! = \! 0$ and $I_t \! = \! J_t$ rules follow in
Sec.~\ref{proof}.  The nature of the $1/N_c$ corrections are described
in Sec.~\ref{1Ncorr}, and we make a few comments about
phenomenological applications and conclude in Sec.~\ref{concl}.

\section{Amplitudes in the $t$ Channel} \label{Tamp}

We consider the baryon-meson scattering process $\phi (S_\phi, R_\phi,
I_\phi, Y_\phi) + B (S_B, R_B, I_B, Y_B) \to \phi^\prime
(S_{\phi^\prime}, R_{\phi^\prime}, I_{\phi^\prime}, Y_{\phi^\prime}) +
B^\prime (S_{B^\prime}, R_{B^\prime}, I_{B^\prime}, Y_{B^\prime})$,
where $S$, $R$, $I$, and $Y$ stand, respectively, for the spin, SU(3)
representation, isospin, and hypercharge of the mesons $\phi$ and
$\phi^\prime$ and the baryons $B$ and $B^\prime$.  Primes indicate
final-state quantum numbers.  We take the baryons to lie in the
ground-state band, the arbitrary-$N_c$ analogue of the SU(6) {\bf 56},
whose lowest states ($N$, $\Delta$, $\Sigma$, etc.) in the large $N_c$
limit are stable against strong decay.  The relative angular momenta
between the meson-baryon pairs are denoted by $L$ and $L^\prime$.  As
shown in the original derivation~\cite{MMSU3}, it is convenient to
cross the quantum numbers of the process to consider instead $\phi \!
+ \! \phi^{\prime *} \to B^* \! + \! B^\prime$.  The amplitude is
described in terms of $t$-channel angular momentum $J_t$, SU(3)
representation $R_t$, isospin $I_t$, and hypercharge $Y_t$.  In
addition, multiple copies of $R_t$ may arise in the products $R_\phi
\otimes R_{\phi^\prime}^*$ and $R^*_B \otimes R_{B^\prime}$, and the
quantum numbers defined to lift this degeneracy are labeled,
respectively, by $\gamma_t$ and $\gamma^\prime_t$ (which need not be
equal).  In the physical amplitude, one of course sums coherently over
all allowed $t$-channel quantum numbers.  After a derivation following
the methods of Ref.~\cite{MMSU3}, we obtain the large $N_c$ master
expression for such scattering amplitudes expressed in the $t$
channel:
\begin{eqnarray}
\lefteqn{S_{L L^\prime S_{B^{\vphantom\prime}} S_{B^\prime} J_t J_{tz}
R_t \gamma^{\vphantom\prime}_t \gamma^\prime_t I_t Y_t} =
\delta_{J^{\vphantom\prime}_t J^\prime_t} \,
\delta_{J^{\vphantom\prime}_{tz} J^\prime_{tz}}
\delta_{R^{\vphantom\prime}_t R^\prime_t} \,
\delta_{I^{\vphantom\prime}_t I^\prime_t} \,
\delta_{I^{\vphantom\prime}_{tz} I^\prime_{tz}}
\delta_{Y^{\vphantom\prime}_t Y^\prime_t} } \nonumber \\
& \times & (-1)^{S_{\phi^\prime} - S_{\phi^{\vphantom\prime}} +
J_\phi - J_t}
([R_{B^{\vphantom\prime}}][R_{B^\prime}]
[J_{\phi^{\vphantom\prime}}] [J_{\phi^\prime}])^{1/2} / [R_t]
\nonumber \\ & \times &
\sum_{\stackrel{\scriptstyle I \in R_{\phi^{\vphantom\prime}}, \,
I^\prime \in R_{\phi^\prime} \! ,}{Y \in R_{\phi^{\vphantom\prime}}
\cap \, R_{\phi^\prime}}} 
\left( \begin{array}{cc||c} R_{\phi^{\vphantom\prime}} &
R^*_{\phi^\prime} & R_t \, \gamma_t \\ I Y & I^\prime \! , \!
- \! Y & J_t \, 0 \end{array} \right)
\left( \begin{array}{cc||c} R_{\phi^{\vphantom\prime}} &
R^*_{\phi^\prime} & R_t \, \gamma_t \\ I_\phi Y_\phi
& I_{\phi^\prime}, \! - \! Y_{\phi^\prime} & I_t \, Y_t
\end{array} \right) \nonumber \\ & \times &
\left( \begin{array}{cc||c} R^*_B & R_{B^\prime} & R_t \,
\gamma^\prime_t \\ S_B, - \frac{N_c}{3} & S_{B^\prime} \! ,
+ \frac{N_c}{3} & J_t \, 0 \end{array} \right)
\left( \begin{array}{cc||c} R^*_B & R_{B^\prime} & R_t \,
\gamma^\prime_t \\ I_B, - \! Y_B & I_{B^\prime} Y_{B^\prime} & I_t \,
Y_t \end{array} \right)
\nonumber \\ & \times & \sum_{K \tilde{K} \tilde{K}^\prime}
(-1)^{K - \frac{Y}{2}} [K] ( [\tilde{K}] [\tilde{K}^\prime] )^{1/2}
\left\{ \begin{array}{ccc}
J_\phi   & I              & K \\
I^\prime & J_{\phi^\prime} & J_t \end{array} \right\} \!
\left\{ \begin{array}{ccc}
J_\phi    & I      & K \\
\tilde{K} & S_\phi & L \end{array} \right\} \!
\left\{ \begin{array}{ccc}
J_{\phi^\prime}  & I^\prime        & K \\
\tilde{K}^\prime & S_{\phi^\prime} & L^\prime \end{array} \right\} \!
\nonumber \\ & \times &
\tau^{\left\{ I I^\prime Y \right\}}_{K \tilde{K} \tilde{K}^\prime L
L^\prime} \ .
\label{tchannel}
\end{eqnarray}
The quantities containing double vertical bars are SU(3) isoscalar
Clebsch-Gordan coefficients (CGC)~\cite{CLSU3}, while those in braces
are ordinary SU(2) $6j$ symbols.  The notation $[X]$ refers to the
dimension of a given representation, whether $X$ is labeled by $I$ or
$J$ in SU(2), or by the actual dimension in SU(3) (i.e., [$J \!  = \!
1$] = 3, but [$R \! = \! {\bf 8}$] = 8).  The quantities $\tau$ are
the reduced amplitudes, which represent the independent dynamical
degrees of freedom in the large $N_c$ limit; in the $1/N_c$ expansion,
Eq.~(1) is corrected both by including $O(1/N_c)$ corrections to the
$\tau$'s, as well as by adding (as discussed in Sec.~\ref{1Ncorr})
terms with group-theoretical structures distinct from those in
Eq.~(\ref{tchannel}) times additional $O(1/N_c)$ reduced amplitudes.

Equation~(\ref{tchannel}) should be compared to the original result
Eq.~(8) of Ref.~\cite{MMSU2} or Eq.~(15) of Ref.~\cite{MMSU3} (the
latter of which provides details of the original derivation).  We
previously showed~\cite{CLSU3penta} in rederiving the corresponding
$s$-channel expressions (Eq.~(7) of~\cite{MMSU2} or Eq.~(12)
of~\cite{MMSU3}) that small but significant discrepancies arise, and
the same comments hold for our rederivation of the $t$-channel
results: First, Ref.~\cite{MMSU3} appears to average over baryons and
mesons in the external states with all possible quantum numbers within
the given SU(3) multiplets; if we do the same with
Eq.~(\ref{tchannel}), two of our SU(3) CGC are absorbed through an
orthogonality relation (Eq.~(11.3$a$) of Ref.~\cite{deSwart}),
matching the form of the older result.  Second, their explicit unity
values for the nonstrange baryon hypercharges must be modified to
$+\frac{N_c}{3}$, in light of the proper quantization~\cite{WZ} of the
Wess-Zumino term for arbitrary $N_c$; similarly, the baryon
representations must be generalized to their proper arbitrary-$N_c$
forms: For example, the literal SU(3) {\bf 8}, becomes an ``{\bf 8}''
= $[1, (N_c \! - \! 1)/2 ]$.  Finally, we obtain a phase quite
different~\cite{phase} from the one in the original result.

The only $N_c$-dependent factors in Eq.~(\ref{tchannel}) appear in the
last two SU(3) CGC and the dimension factors $[R_B],[R_{B^\prime}]$,
which refer to the large baryon representations.  Focusing only on
these factors, one may use SU(3) CGC reflection
properties~(Eqs.~(14.9) and (14.13) of Ref.~\cite{deSwart}), augmented
by a proper treatment of phase factors~\cite{phase} for
representations with non-integer hypercharges, and Eq.~(1) of
Ref.~\cite{CL70} ($[R_B] \to N_c^2 [S_B] /8$):
\begin{eqnarray}
& & \frac{\sqrt{[R_{B^{\vphantom\prime}}][R_{B^\prime}]}}{[R_t]}
\left( \begin{array}{cc||c} R^*_B & R_{B^\prime} & R_t \,
\gamma^\prime_t \\ S_B, - \frac{N_c}{3} & S_{B^\prime} \! ,
+ \frac{N_c}{3} & J_t \, 0 \end{array} \right)
\left( \begin{array}{cc||c} R^*_B & R_{B^\prime} & R_t \,
\gamma^\prime_t \\ I_B, - \! Y_B & I_{B^\prime} Y_{B^\prime} & I_t \,
Y_t \end{array} \right)
\nonumber \\ & = &
(-1)^{(I_{B^\prime} - S_{B^\prime}) -\frac 1 2 (Y_B - \frac{N_c}{3}) -
(I_t - J_t)} \nonumber \\ & \times &
\sqrt{\frac{[S_B][I_{B^\prime}]}{[I_t][J_t]}}
\left( \begin{array}{cc||c} R_B & R_t & R_{B^\prime} \, \tilde \gamma
\\ S_B, + \frac{N_c}{3} & J_t \, 0 & S_{B^\prime} \! , + \frac{N_c}{3}
\end{array} \right)
\left( \begin{array}{cc||c} R_B & R_t & R_{B^\prime} \, \tilde \gamma
\\ I_B Y_B & I_t \, Y_t & I_{B^\prime} Y_{B^\prime}
\end{array} \right) \ . \label{reflect}
\end{eqnarray}
Since the baryons with $N_s$ strange quarks have $Y_B \! = \!
\frac{N_c}{3} \! - \! N_s$, all $N_c$-dependent factors are
relegated to the two new CGC.  Again, the complete amplitude requires
a coherent sum over multiplicity factors, in this case $\tilde
\gamma$.  We therefore seek to prove that, at $O(N_c^0)$,
\begin{equation}
\sum_{\tilde \gamma}
\left( \begin{array}{cc||c} R_B & R_t & R_{B^\prime} \, \tilde \gamma
\\ S_B, + \frac{N_c}{3} & J_t \, 0 & S_{B^\prime} \! , + \frac{N_c}{3}
\end{array} \right)
\left( \begin{array}{cc||c} R_B & R_t & R_{B^\prime} \, \tilde \gamma
\\ I_B Y_B & I_t \, Y_t & I_{B^\prime} Y_{B^\prime}
\end{array} \right) \propto \delta_{I_t J_t} \ . \label{ItJtEquiv}
\end{equation}

\section{Proving the $I_t \! = \! J_t$ Rule} \label{proof}

\subsection{The $Y_t \! = \! 0$ Rule} \label{Yt0}

We begin by recalling the theorem demonstrated in Ref.~\cite{CL70}:
Let $R_B \!  = \! (2S_B, \, \frac{N_c}{2} \! - \! S_B)$ denote an
SU(3) representation corresponding to baryons in the ground-state
SU(6) ``{\bf 56}'' with spin $S_B$, so that the top (nonstrange) row
in the weight diagram has isospin $I_{B, \rm top} \!  = \! S_B$ and
$Y_{B, \rm max} \! = \! +\frac{N_c}{3}$, let $R_\phi \!  = \!
(p_\phi, \, q_\phi)$ be an SU(3) (meson) representation with weights
$p_\phi$, $q_\phi \!  = \!  O(N_c^0)$, and let $R_s \gamma_s \subset
R_B \otimes R_\phi$, where $Y_{s, \rm max} \! = \! \frac{N_c}{3} + r$
and $R_s = (2I_{s, \rm top}, \, \frac{N_c}{2} + \frac{3r}{2} - I_{s,
\rm top})$, $r \! = \!  O(N_c^0)$.  Then the SU(3) CGC satisfy
\begin{equation} \label{CGCmag}
\left( \begin{array}{cc||c}
R_B & R_\phi & R_s \, \gamma_s \\ I_B, \frac{N_c}{3} \! - \! m &
I_\phi Y_\phi & I_s, \, \frac{N_c}{3} \! + \! Y_\phi \! - \! m
\end{array} \right) \le O(N_c^{-|Y_\phi - r|/2}) \ ,
\end{equation}
for all allowed $O(N_c^0)$ values of $m$, saturation of the inequality
occurring for almost all CGC\@.  In words: The $O(N_c^0)$ CGC must
have a meson hypercharge that equals the hypercharge difference
between the tops of the two baryon representations.

The CGC in Eq.~(\ref{ItJtEquiv}) have $R_s \! = \! R_{B^\prime}$,
which lies in the ground-state ``{\bf 56}'', so that $r \! = \! 0$.
The first CGC in Eq.~(\ref{ItJtEquiv}) is thus automatically of
leading [$O(N_c^0)$] order, while the second is of leading order only
for $Y_t \! = \! 0$.  This result, derived using the same theorem
Eq.~(\ref{CGCmag}), was noted (in $s$-channel language) in
Ref.~\cite{CL70}; since here we use explicit $t$-channel expressions,
we call it the $Y_t \! = \! 0$ {\it rule}.

At the level of quark diagrams and combinatorics, the $Y_t \! = \! 0$
rule is perfectly sensible.  Baryon-meson scattering diagrams all
involve gluon and/or quark exchanges; the latter diagrams combine to
produce the proper $O(N_c^0)$ amplitude only when each of the $N_c$
quarks in the baryon is permitted to be the one that is exchanged with
the meson.  Furthermore, strangeness-changing ($Y_t \! \neq \!  0$)
baryon-meson scattering can only proceed through such a (strange)
quark exchange.  Since the large-$N_c$ counterparts of the physical
baryons possess only $O(N_c^0)$ strange quarks, such $Y_t \! \neq \!
0$ processes require an exchange of the comparatively rare baryon $s$
quarks, costing a factor of $N_c^{1/2}$ in the amplitude (once the
proper wave function normalization is taken into account).  Thus, at
$O(N_c^0)$ one has $Y_t \! = \! 0$.

Alternately, using the familiar operator approach to baryonic matrix
elements~\cite{operator}, one may observe that strangeness-changing
operators with $O(N_c^{1/2})$ matrix elements do indeed occur (such as
those of the combined spin- ($i$) flavor ($a$) operator $G^{ia}$ with
$a \! = \!  4,5,6,7$) when sandwiched between two states with
$O(N_c^0)$ strange quarks.

\subsection{$I_t \! = \! J_t$: Nonstrange Case}

We already possess from Ref.~\cite{CL70} the elements of a proof that
the $I_t \! = \! J_t$ rule holds for nonstrange baryon-meson
scattering; in \cite{CL70} we showed that the 3-flavor scattering
amplitude expressed in the $s$ channel reduces for nonstrange
processes to the long-known 2-flavor result~\cite{MP}.  This, in turn,
was the equation that MM used to prove~\cite{MMSU2} the $I_t \! = \!
J_t$ rule.  Since the 3-flavor $s$-channel and $t$-channel results are
necessarily equivalent---they obtain from the same source and use the
same formalism---the $I_t \! = \! J_t$ rule must directly follow as a
result of the $t$-channel result restricted to 2 flavors.  Notably,
however, the authors of Ref.~\cite{MMSU2} state their inability to
prove this step.

In fact, the missing ingredients in Ref.~\cite{MMSU2} are precisely
those described in the last section, that $N_c$-dependent factors such
as the sizes of baryon representations and their hypercharges must be
treated correctly.  Here we show that the nonstrange $I_t \!  = \!
J_t$ rule follows directly from the 3-flavor $t$-channel expression
Eqs.~(\ref{tchannel})--(\ref{reflect}).  Our previous $s$-channel
proof~\cite{CL70} mandates this result, but it is not merely
instructional to prove the result using the $t$-channel expression: As
we see in the next subsection, this exercise provides the necessary
impetus to prove $I_t \! = \! J_t$ in the 3-flavor case.  But first,
the nonstrange case:

Consider only the two SU(3) CGC appearing in Eq.~(\ref{ItJtEquiv}).
We have already proved that $Y_t \! = \! 0$ for leading-order
processes.  For the nonstrange case, $Y_B \! = \! +\frac{N_c}{3} \! =
\!  Y_{B^\prime}$, which specifies states in the singly-degenerate top
row of their respective SU(3) multiplets, $R_B$ and $R_{B^\prime}$,
within the ground-state ``{\bf 56}''.  In particular, knowing
$S_{B^\prime}$ uniquely specifies $R_{B^\prime} \! = \! (
S_{B^\prime}, \frac{N_c}{2} \! - \! S_{B^\prime} )$ among the ``{\bf
56}'' states.

Using Eq.~(\ref{CGCmag}) however, one can turn the argument around and
show that $R_{B^\prime}$ can {\em only\/} lie in the ``{\bf 56}''.  In
the context of Eq.~(\ref{CGCmag}), we have $I_B \! = \! S_B$, $m \! =
\!  0$, and $Y_\phi \! = \! 0$, meaning that the $O(N_c^0)$ CGC all
have $r \!  = \! 0$ and thus $Y_{s, \rm max} \! = \! +\frac{N_c}{3}$
and $R_{B^\prime} \! = \! R_s \! = \! (2I_{s, \rm top}, \,
\frac{N_c}{2} - I_{s, \rm top})$, which are precisely the SU(3)
representations lying in ``{\bf 56}'', and therefore $I_{B^\prime} \!
= \! I_{s, \rm top} \!  = \!  S_{B^\prime}$.  The combination of these
two observations tells us that choosing $S_{B^\prime}$ specifies one
and only one $R_{B^\prime}$; therefore, one may sum over
$R_{B^\prime}$ without changing Eq.~(\ref{ItJtEquiv}).  Moreover, the
hypercharges in the kets of Eq.~(\ref{ItJtEquiv}) are fixed to equal
$+\frac{N_c}{3}$, but this entry may be replaced with a variable
$\tilde Y$ and summed over without loss of generality; the CGC of
Eq.~(\ref{ItJtEquiv}) may thus be replaced by
\begin{equation} \label{ItJtNS}
\sum_{R_{B^\prime}, \tilde \gamma, \tilde Y}
\left( \begin{array}{cc||c} R_B & R_t & R_{B^\prime} \, \tilde \gamma
\\ S_B, + \frac{N_c}{3} & J_t \, 0 & S_{B^\prime} \! , \tilde Y
\end{array} \right)
\left( \begin{array}{cc||c} R_B & R_t & R_{B^\prime} \, \tilde \gamma
\\ S_B, +\frac{N_c}{3} & I_t \, 0 & S_{B^\prime} \! , \tilde Y
\end{array} \right) = \delta_{I_t J_t} \ .
\end{equation}
The final equality, which is precisely the $I_t \! = \! J_t$ rule, is
just a special case of the orthogonality relation Eq.~(11.3$b$) of
Ref.~\cite{deSwart}.  In light of our previous comments, the sums over
$R_{B^\prime}$ and $\tilde Y$ are unnecessary, but the sum over
$\tilde \gamma$ is required; one may check this explicitly in cases
where the relevant CGC are tabulated~\cite{CLSU3}.

\subsection{$I_t \! = \! J_t$: The 3-Flavor Case}

Now we return to the CGC in Eq.~(\ref{ItJtEquiv}), but note that $Y_t
\!  = \! 0$ still holds, so that $Y_{B^\prime} \! = \! Y_B$.  The
second, but not the first, CGC in (\ref{ItJtEquiv}) depends upon
quantum numbers corresponding to nonzero strangeness.  We claim that
one may obtain the second CGC from the first by means of repeated
applications of recursion relations, and that no step in this
recursion depends upon the value of $\tilde \gamma$.  It then follows
that the second CGC is proportional to the first, with the
proportionality constant being a complicated function of all the
quantum numbers in the two CGC except for $\tilde \gamma$.  But then,
these complicated prefactors simply pull through the sum on $\tilde
\gamma$, and the sum reduces to the one that we obtained in
Eq.~(\ref{ItJtNS}).

Proving the 3-flavor $I_t \! = \! J_t$ rule therefore requires one
only to one show that the first CGC in Eq.~(\ref{ItJtEquiv}) uniquely
sets the scale for all CGC of the form of the second CGC in
(\ref{ItJtEquiv}), independent of $\tilde \gamma$.  Of course, their
absolute sizes are determined by unitarity.

The required recursion relations are none other than those for the
strangeness-changing SU(3) ladder operators $U_{\pm}$ and $V_{\pm}$,
which indeed were the ingredients used to prove Eq.~(\ref{CGCmag}).
In general, such relations involve six SU(3) CGC (e.g., Eq.~(2.5) of
Ref.~\cite{CLSU3}).  However, in the case of large $N_c$ baryon-meson
couplings, the large square-root prefactors (analogues to the familiar
SU(2) factors $[(I \mp I_z) (I \pm I_z +1)]^{1/2}$ associated with
operators $I_\pm$) that change meson hypercharge (and isospin) are
relatively smaller by a factor $N_c^{-1/2}$~\cite{CL70}.
Incidentally, these suppressed factors are also the only ones that
depend upon $R_t$.

Therefore, for large $N_c$, recursion relations based upon $U_{\pm}$
and $V_{\pm}$ connect only CGC with a fixed $I_t$ and $Y_t \! = \!
0$.  Now it remains only to show that these recursion relations point
uniquely back to the nonstrange CGC in Eq.~(\ref{ItJtEquiv}),
\begin{equation} \label{CGCtop}
\left( \begin{array}{cc||c} R_B & R_t & R_{B^\prime} \, \tilde \gamma
\\ S_B, +\frac{N_c}{3} & I_t \, 0 & S_{B^\prime}, +\frac{N_c}{3}
\end{array} \right) .
\end{equation}
But this is not difficult to show, for consider an arbitrary CGC
having the form of the second one in Eq.~(\ref{ItJtEquiv}), with $Y_t
\! = \!  0$.  One may repeatedly apply recursion relations that
increase $Y_B \! = \! Y_{B^\prime}$ by one unit at each step until one
reaches $Y_{B, \rm top} \! = \!  +\frac{N_c}{3}$.  However, the top
row of the baryon weight diagrams consists of singly-occupied sites of
a unique isospin, and consequently the only CGC appearing at that
level is the one given by Eq.~(\ref{CGCtop}).  In most cases, this
completes the proof.

But one special exception must be noted.  Given an arbitrary allowed
CGC having the form of the second one in Eq.~(\ref{ItJtEquiv}), it can
occur that the given value of $I_t$ satisfies the triangle rule
$\delta ( I_B \, I_t \, I_{B^\prime} )$ but not the nonstrange
triangle rule $\delta ( S_B \, I_t \, S_{B^\prime} )$, and hence the
CGC Eq.~(\ref{CGCtop}) vanishes.  Since $J_t$ by construction
satisfies the triangle rule $\delta ( S_B \, J_t \, S_{B^\prime} )$,
it follows that this particular value of $I_t$ cannot equal $J_t$.
But how does one then prove that the expression in
Eq.~(\ref{ItJtEquiv}) vanishes?  In that case, one simply notes that
the recursion process upwards in values of $Y_B$ leads eventually to a
value of hypercharge where the quantum numbers are no longer allowed,
and the CGC vanishes.  But one may then reverse the process, applying
hypercharge-lowering operators to such a {\em disallowed\/} CGC to
obtain recursion relations for nominally {\em allowed\/} CGC of lower
hypercharge, including the ones we start with.  All such CGC must
therefore vanish for large $N_c$ (but might survive for finite $N_c$),
guaranteeing that Eq.~(\ref{ItJtEquiv}) vanishes.  This mechanism,
incidentally, is the origin of such CGC that are nonzero but do not
saturate the bound given by Eq.~(\ref{CGCmag}).

In light of the $Y_t \! = \! 0$ and $I_t \! = \! J_t$ rules, and using
Eq.~(\ref{reflect}), the master expression Eq.~(\ref{tchannel}) is
most conveniently written (keeping factors originating as $I_t$ or
$J_t$ distinct) as
\begin{eqnarray}
\lefteqn{S_{L L^\prime S_{B^{\vphantom\prime}} S_{B^\prime} J_t J_{tz}
R_t \gamma^{\vphantom\prime}_t \gamma^\prime_t I_t Y_t} =
\delta_{J^{\vphantom\prime}_t J^\prime_t} \,
\delta_{J^{\vphantom\prime}_{tz} J^\prime_{tz}}
\delta_{R^{\vphantom\prime}_t R^\prime_t} \,
\delta_{I^{\vphantom\prime}_t I^\prime_t} \,
\delta_{I^{\vphantom\prime}_{tz} I^\prime_{tz}}
\delta_{Y^{\vphantom\prime}_t Y^\prime_t}
\delta_{I_t J_t} \delta_{Y_t, 0} }
\nonumber \\
& \times & (-1)^{S_{\phi^\prime} - S_{\phi^{\vphantom\prime}} +
J_\phi - J_t + (I_{B^\prime} - S_{B^\prime})
-\frac 1 2 (Y_B - \frac{N_c}{3})}
([S_B][I_{B^\prime}]
[J_{\phi^{\vphantom\prime}}] [J_{\phi^\prime}]/[I_t][J_t])^{1/2}
\nonumber \\ & \times &
\sum_{\stackrel{\scriptstyle I \in R_{\phi^{\vphantom\prime}}, \,
I^\prime \in R_{\phi^\prime} \! ,}{Y \in R_{\phi^{\vphantom\prime}}
\cap \, R_{\phi^\prime}}} 
\left( \begin{array}{cc||c} R_{\phi^{\vphantom\prime}} &
R^*_{\phi^\prime} & R_t \, \gamma_t \\ I Y & I^\prime \! , \!
- \! Y & J_t \, 0 \end{array} \right)
\left( \begin{array}{cc||c} R_{\phi^{\vphantom\prime}} &
R^*_{\phi^\prime} & R_t \, \gamma_t \\ I_\phi Y_\phi
& I_{\phi^\prime}, \! - \! Y_{\phi^\prime} & I_t \, 0
\end{array} \right) \nonumber \\ & \times &
\left( \begin{array}{cc||c} R_B & R_t & R_{B^\prime} \, \tilde \gamma
\\ S_B, + \frac{N_c}{3} & J_t \, 0 & S_{B^\prime} \! , + \frac{N_c}{3}
\end{array} \right)
\left( \begin{array}{cc||c} R_B & R_t & R_{B^\prime} \, \tilde \gamma
\\ I_B Y_B & I_t \, 0 & I_{B^\prime} Y_B
\end{array} \right)
\nonumber \\ & \times & \sum_{K \tilde{K} \tilde{K}^\prime}
(-1)^{K - \frac{Y}{2}} [K] ( [\tilde{K}] [\tilde{K}^\prime] )^{1/2}
\left\{ \begin{array}{ccc}
J_\phi   & I              & K \\
I^\prime & J_{\phi^\prime} & J_t \end{array} \right\} \!
\left\{ \begin{array}{ccc}
J_\phi    & I      & K \\
\tilde{K} & S_\phi & L \end{array} \right\} \!
\left\{ \begin{array}{ccc}
J_{\phi^\prime}  & I^\prime        & K \\
\tilde{K}^\prime & S_{\phi^\prime} & L^\prime \end{array} \right\} \!
\nonumber \\ & \times &
\tau^{\left\{ I I^\prime Y \right\}}_{K \tilde{K} \tilde{K}^\prime L
L^\prime} \ .
\end{eqnarray}

\section{$1/N_c$ Corrections} \label{1Ncorr}

The arguments of Refs.~\cite{KapSavMan} that demonstrate the 2-flavor
$I_t \! = \! J_t$ rule in baryon-baryon scattering using the operator
approach~\cite{operator} can be generalized, not only to the
baryon-meson case~\cite{CL1st}, but to three flavors and to
delineating the form of $1/N_c$ corrections as well.

The primary tool in the 2-flavor case is the observation that
arbitrary $n$-body operators (i.e., having $n$ quark creation and
destruction operators) can be written in terms of products of $n$
1-body operators, and give matrix elements subleading in the $1/N_c$
expansion unless all of the 1-body operators are either of the form
$\openone$ (quark number operators) or $G^{ia}$.  Then, the operator
reduction rules~\cite{operator} indicate that contractions of indices
among the $G$'s always lead to operators with matrix elements of lower
order than $O(N_c^1)$ for each $G$, while uncontracted $G$'s may be
symmetrized among their spin and flavor indices---one of each for
every $G$---and therefore the leading-order operator has $I_t \! = \!
J_t$.  Each contraction or non-$G$, non-$\openone$ 1-body operator
(i.e., pure isospin $I^a$ or spin $J^i$) costs a relative factor
$N_c$, and therefore operators with $|I_t \! - \! J_t| \! = \! n$ are
suppressed by a relative factor $1/N_c^n$~\cite{CL1N,CLphoto}.

These arguments may be generalized to three flavors (as, indeed, was
strongly suggested in Refs.~\cite{KapSavMan}).  The only additional
1-body operator with $O(N_c^1)$ matrix elements on the baryons with
$N_s \! = \! O(N_c^0)$ baryons is $T^8$, but its $O(N_c^1)$ part is
simply proportional to $\openone$.  Next, in principle the factors of
$G^{ia}$ with $a \! = \! 4,5,6,7,8$ (i.e., not isovector) spoil the
$I_t \! = \! J_t$ rule.  However, as observed in Subsec.~\ref{Yt0},
the strangeness-changing components of $G$ give only $O(N_c^{1/2})$
matrix elements on these baryon states, as do the strangeness-changing
matrix elements of $T^a$.  Finally, $G^{i8}$ on these states has only
$O(N_c^0)$ matrix elements~\cite{operator}.  Thus, only the $I_t \! =
\! J_t$ portions of $G^{ia}$ ($a \! = \! 1,2,3$) contribute to the
leading-order amplitudes; in fact, this can be taken as an alternate
proof of our primary conclusion.

But this argument also indicates the nature of the $1/N_c$
corrections.  For the $N_s \! = \! O(N_c^0)$ states, each
strangeness-changing operator costs a factor of $N_c^{1/2}$, while
each unit of $|I_t \! - \! J_t|$, by the same arguments as before,
costs a factor of $N_c$.  The $Y_t \! = \! 0$ rule has an additional
$O(N_c^{-1/2})$ correction associated with each unit of strangeness
change, while within a sector of fixed $Y_t \! = \! 0$, the $I_t \! =
\! J_t$ rule has an additional $O(1/N_c)$ correction for each unit of
difference between $I_t$ and $J_t$.

\section{Conclusions} \label{concl}

We have shown that baryon-meson scattering amplitudes, regardless of
the strangeness content of the hadrons involved, satisfy two selection
rules when expressed as $t$-channel exchanges: $Y_t \! = \! 0$ and
$I_t \! = \! J_t$.  We have also explained how to characterize their
$1/N_c$ corrections by means of their quantum numbers:
$N_c^{-|Y_t|/2}$ and $N_c^{-|I_t \! - \! J_t|}$, respectively.

The phenomenological implications are immediate, but their detailed
application will be reserved for another paper.  For example, the
process $K^- p \to \pi^+ \Sigma^-$ is suppressed in cross section by
$1/N_c$ compared to, say, $K^- p \to K^- p$.  More restrictive,
however, will be relations among amplitudes with no strangeness
exchange, such as $KN \to KN$.  Such constraints were implicitly used
in studying possible pentaquark multiplets~\cite{CLpenta,CLSU3penta},
and indeed can be considered as relations not relying on perfect SU(3)
symmetry but only SU(2)$\times$U(1) symmetry along a line of fixed
strangeness.  Clearly, an SU(3)-derived selection rule that does not
require SU(3) symmetry will provide robust information.

{\it Acknowledgments.}  I thank Tom Cohen for enlightening comments
and the very successful collaboration that made this project possible.
This work supported in part by the N.S.F.\ through grant No.\
PHY-0456520.

\end{document}